\documentclass[pra,twocolumn,preprintnumbers,showpacs,nofootinbib]{revtex4}
\usepackage{amsmath,amssymb,bm,graphicx}

\renewcommand\d{\partial}
\newcommand\grad{\bm{\nabla}}
\newcommand\+{\dagger}
\newcommand\<{\langle}
\renewcommand\>{\rangle}
\renewcommand\r{{\bm{r}}}
\newcommand\x{{\bm{x}}}
\newcommand\y{{\bm{y}}}
\newcommand\p{{\bm{p}}}
\newcommand\q{{\bm{q}}}
\renewcommand\k{{\bm{k}}}
\newcommand\kF{k_\mathrm{F}}
\newcommand\nF{n_{\mathrm{F}}}
\newcommand\Tc{T_{\mathrm{c}}}

\begin{document}
\preprint{MIT-CTP 3996}

\title{Casimir interaction among heavy fermions in the BCS-BEC crossover}
\author{Yusuke~Nishida}
\affiliation{Center for Theoretical Physics, Massachusetts
Institute of Technology, Cambridge, Massachusetts 02139, USA}

\begin{abstract}
 We investigate a two-species Fermi gas with a large mass ratio
 interacting by an interspecies short-range interaction.  Using the
 Born-Oppenheimer approximation, we determine the interaction energy of
 two heavy fermions immersed in the Fermi sea of light fermions as a
 function of the $s$-wave scattering length.  In the BCS limit, we
 recover the perturbative calculation of the effective interaction
 between heavy fermions.  The $p$-wave projection of the effective
 interaction is attractive in the BCS limit while it turns out to be
 repulsive near the unitarity limit.  We find that the $p$-wave
 attraction reaches its maximum between the BCS and unitarity limits,
 where the maximal $p$-wave pairing of heavy minority fermions is
 expected.  We also investigate the case where the heavy fermions are
 confined in two dimensions and the $p$-wave attraction between them is
 found to be stronger than that in three dimensions.
\end{abstract}

\date{November 2008}
\pacs{03.75.Ss, 05.30.Fk, 67.85.Lm}

\maketitle

\section{Introduction}
Experiments using ultracold atomic gases have achieved great success in
realizing a new type of fermionic superfluid.  By arbitrarily varying
the strength of interaction via the Feshbach resonance, the crossover
from the BCS superfluid of fermionic atoms to the Bose-Einstein
condensate of tightly bound molecules has been observed and extensively
studied~\cite{Ketterle:2008,review_theory}.  Nowadays a portion of
interests of the cold atom community is shifting to the two-species
Fermi gas with unequal
densities~\cite{Zwierlein:2005,Partridge:2005,Zwierlein:2006,Shin:2006,Partridge:2006}
and with unequal masses~\cite{Taglieber:2008,Wille:2008,Voigt:2008}.
Elucidating the phase diagram of such an asymmetric Fermi gas is an
important and challenging problem.  Such an asymmetric system of
fermions will be also interesting as a prototype of high density quark
matter in the core of neutron stars, where the density and mass
imbalances exist among different quark flavors~\cite{Alford:2007xm}.

So far several quantum phases have been proposed as a ground state of
the density-imbalanced Fermi gas.   One of such phases is the $p$-wave
superfluid phase by the pairing between the same species of
fermions~\cite{Bulgac:2006gh}.  In Ref.~\cite{Bulgac:2006gh}, the
attraction between the same species of fermions was found to be induced
by the interaction with the other species of fermions based on the
perturbative calculation in the weak-coupling BCS limit.  Although the
resulting pairing gap is exponentially suppressed in the BCS limit, it
may be possible that the pairing gap becomes large enough near the
strong-coupling unitarity limit.  Thus an important question is how
large the pairing gap can be near the unitarity limit.

However, difficulties for theoretical treatments arise away from the
weak-coupling BCS limit because we do not have a controlled tool to
analyze the strongly interacting many-body system.  Monte Carlo
simulations also have limitations in asymmetric systems due to the
fermion sign problem.  We note that in one dimension the Monte Carlo
simulations do not suffer from the sign problem and have been employed
to study the Fulde-Ferrel-Larkin-Ovchinikov phase in the
density-imbalanced Fermi
gas~\cite{Batrouni:2008,Casula:2008,Batrouni:2009}.

In order to obtain further insight into the density-imbalanced Fermi gas
and the $p$-wave pairing therein, we investigate the Fermi gas with
unequal masses between two different atomic species.  In the limit of
large mass ratio, we can determine the effective interaction among heavy
fermions induced by the interaction with light fermions using the
Born-Oppenheimer approximation.  Because we do not need to rely on the
weak-coupling approximation, we can go beyond the perturbative BCS
regime to the strongly interacting unitary regime in a controlled way.

We will show that the $p$-wave projection of the effective interaction
between two heavy fermions is attractive in the BCS limit being
consistent with the perturbative prediction, while it turns out to be
repulsive near the unitarity limit.  Thus the $p$-wave attraction has
a maximum between the BCS and unitarity limits, where the maximal
$p$-wave pairing of heavy minority fermions is expected.  We note that
our results have a direct relevance to the recently realized Fermi-Fermi
mixture of ${}^{40}\mathrm{K}$ and ${}^6\mathrm{Li}$ because of their
large mass ratio~\cite{Taglieber:2008,Wille:2008,Voigt:2008}.

It is worthwhile to point it out that the effective interaction among
heavy fermions immersed in the Fermi sea of light fermions is an analog
of the Casimir force among objects placed in a
vacuum~\cite{Casimir:1948dh}.  The origin of the Casimir force can be
traced back to the modification of the spectrum of zero point
fluctuations of the electromagnetic field.  In our case, the role of
vacuum is played by the Fermi sea of light fermions whose spectrum is
modified by the presence of heavy fermions~\cite{Bulgac:2001np}.

This paper is organized as follows.  In Sec.~\ref{sec:casimir}, we
determine the effective interaction between two heavy fermions induced
by the interaction with the Fermi sea of light fermions as a function of
the $s$-wave scattering length using the Born-Oppenheimer approximation.
(The effective interaction for a general number of heavy fermions is
derived using the functional integral method and evaluated in
Appendix~\ref{sec:functional}.)  With having in mind the application to
the $p$-wave pairing of heavy fermions, we compute the $p$-wave
projection of the effective interaction in Sec.~\ref{sec:p-wave}.  Here
we consider both cases where the heavy fermions are in three dimensions
(3D) and where they are confined in two dimensions (2D) while light fermions
are always in 3D.  The latter case corresponds to the two-species Fermi
gas in 2D-3D mixed dimensions studied in
Refs.~\cite{Nishida:2008kr,Nishida:2008gk}.  Finally summary and
concluding remarks are given in Sec.~\ref{sec:summary}.

\section{Casimir interaction between two heavy fermions \label{sec:casimir}}

\subsection{Born-Oppenheimer approximation}
Consider a two-species Fermi gas with unequal masses $M$ and $m$
interacting by an interspecies short-range interaction.  When the mass
ratio is large $M/m\gg1$, we can employ the Born-Oppenheimer
approximation to compute the interaction energy of heavy fermions
immersed in the Fermi sea of light fermions.  Here we concentrate on the
case of two heavy fermions.  The formula for a general number of heavy
fermions is derived in Appendix~\ref{sec:functional} using the
functional integral method.  The wave function of a light fermion
interacting with two heavy fermions fixed at positions $\x_1$ and $\x_2$
satisfies the Schr\"odinger equation (here and below $\hbar=1$):
\begin{equation}\label{eq:schrodinger}
 -\frac{\grad_{\!\y}^2}{2m}\,\Psi(\y;\x_1,\x_2) = E\,\Psi(\y;\x_1,\x_2).
\end{equation}
The interspecies short-range interaction between the light and heavy
fermions is taken into account by imposing the short-range boundary
condition;
$\Psi(\y\to\x_i)\propto\frac1{|\y-\x_i|}-\frac1{a}+O(|\y-\x_i|)$, where
$a$ is the $s$-wave scattering length.  Below we determine the energy
spectrum of the light fermion both for bound states and continuum states
as a function of $a$ and the separation between the heavy fermions
$\r\equiv\x_1-\x_2$.

For the bound state $E=-\frac{\kappa^2}{2m}<0$, the solution of
Eq.~(\ref{eq:schrodinger}) takes the form
\begin{equation}
 \Psi_\pm(\y) = \frac{e^{-\kappa|\y-\x_1|}}{|\y-\x_1|}
  \pm \frac{e^{-\kappa|\y-\x_2|}}{|\y-\x_2|},
\end{equation}
where $\pm$ is the parity of the wave function.  The upper (lower) sign
corresponds to the even (odd) parity.  From the short-range boundary
condition, we obtain the equation
\begin{equation}
 -\kappa \pm \frac{e^{-\kappa|\r|}}{|\r|} = -\frac1a.
\end{equation}
The solution to this equation exists when $|\r|/a>\mp1$ and is given by
\begin{equation}
 \kappa_\pm = \frac1a + \frac1{|\r|}\,W\!\left(\pm e^{-|\r|/a}\right),
\end{equation}
where $W(z)$ is the Lambert function satisfying $z=W(z)e^{W(z)}$.
Accordingly we find the following binding energy depending on $a$ and
$|\r|$:
\begin{equation}
 E_\mathrm{binding}^{(\pm)} = -\frac{\kappa_{\pm}^{\,2}}{2m}
\end{equation}
for $|\r|/a>\mp1$.

For the continuum state $E=\frac{k^2}{2m}>0$, the solution of
Eq.~(\ref{eq:schrodinger}) takes the form
\begin{equation}
 \Psi_\pm(\y) = \frac{\sin\left(k|\y-\x_1|+\delta_\pm\right)}{|\y-\x_1|}
  \pm \frac{\sin\left(k|\y-\x_2|+\delta_\pm\right)}{|\y-\x_2|},
\end{equation}
where $0\leq\delta_\pm\leq\pi$ is the phase shift.  From the short-range
boundary condition, we obtain the equation
\begin{equation}
 k\cos\delta_\pm \pm \frac{\sin\left(k|\r|+\delta_\pm\right)}{|\r|}
  = -\frac{\sin\delta_\pm}{a}.
\end{equation}
The solution to this equation is easily found to be
\begin{equation}\label{eq:delta}
 \tan\delta_\pm = -\frac{k|\r|\pm\sin(k|\r|)}{\frac{|\r|}a\pm\cos(k|\r|)}.
\end{equation}
We then suppose that the system is confined in a large sphere with a
radius $R\gg|\r|$ and the two heavy fermions are located near its
center; $\x_1=\frac{\r}2$ and $\x_2=-\frac{\r}2$.  By imposing
$\Psi_\pm(|\y|\to R)\to0$ at the boundary of the sphere, the momentum
$k$ is discretized as
\begin{equation}
 \begin{split}
  k_n^+R+\delta_+ &= n\pi, \\
  k_n^-R+\delta_- &= \left(n-\frac12\right)\pi
 \end{split}
\end{equation}
with $n=1,2,3,\ldots$.  Therefore the energy level becomes
\begin{equation}
 E_n^{(\pm)} = \frac{{k_n^\pm}^2}{2m}.
\end{equation}

\begin{figure*}[tp]\hfill
 \includegraphics[width=0.46\textwidth,clip]{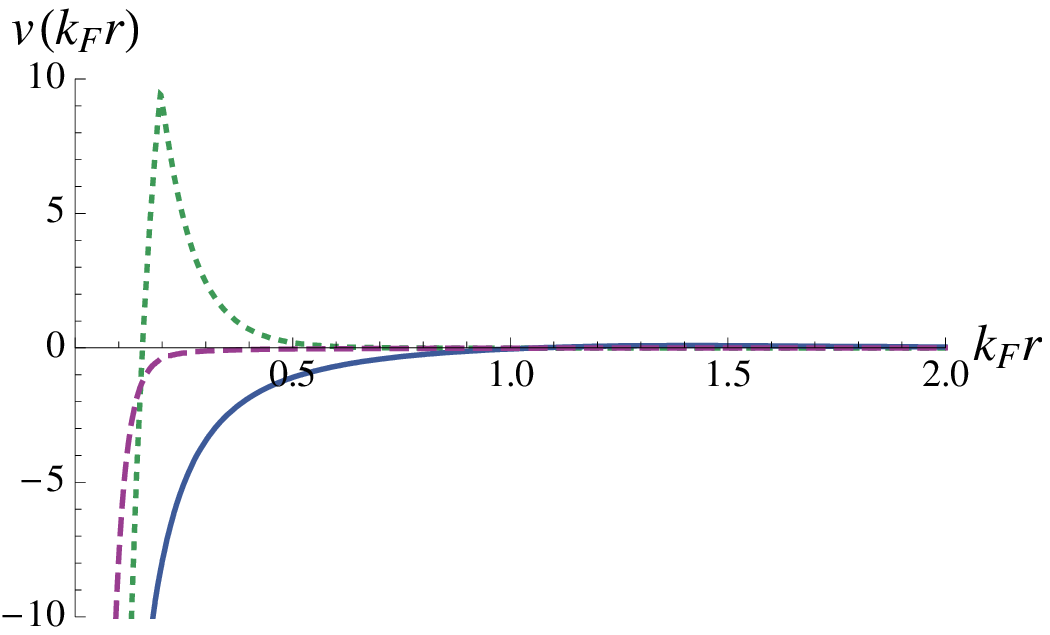}\hfill\hfill
 \includegraphics[width=0.46\textwidth,clip]{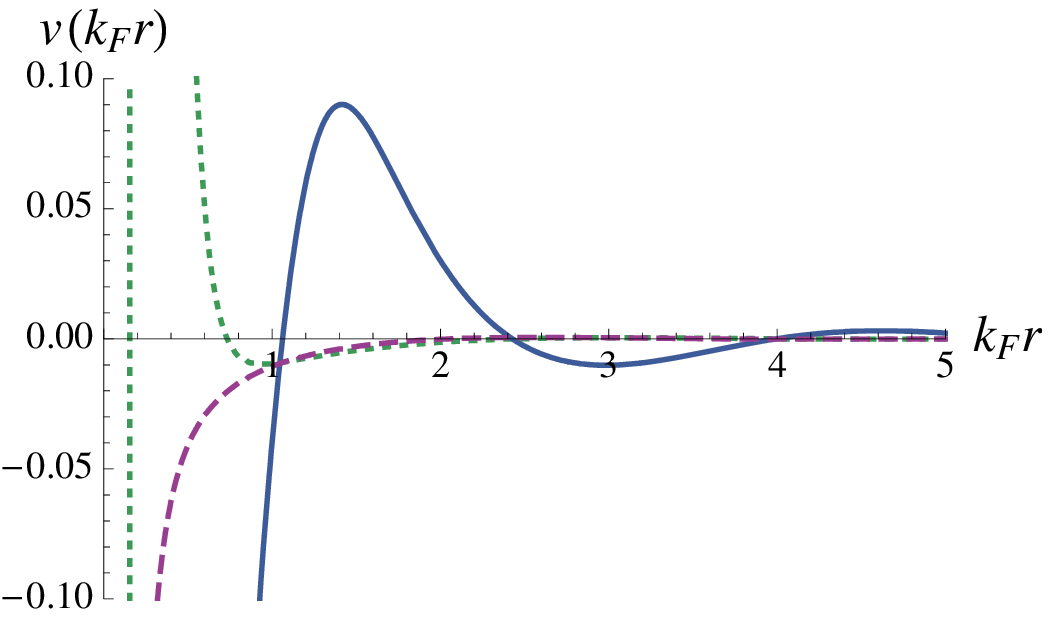}\hfill\hfill
 \caption{(Color online) Interaction energy of two heavy fermions
 $v(\kF r)$ as a function of their separation $\kF r$.  Three different
 curves correspond to the BCS regime $(a\kF)^{-1}=-5$ (dashed curve),
 unitarity limit $(a\kF)^{-1}=0$ (solid curve), and BEC regime
 $(a\kF)^{-1}=5$ (dotted curve).  The right panel is a magnification of
 the left panel.  \label{fig:effective_potential}}
\end{figure*}
\begin{figure*}[tp]\hfill
 \includegraphics[width=0.46\textwidth,clip]{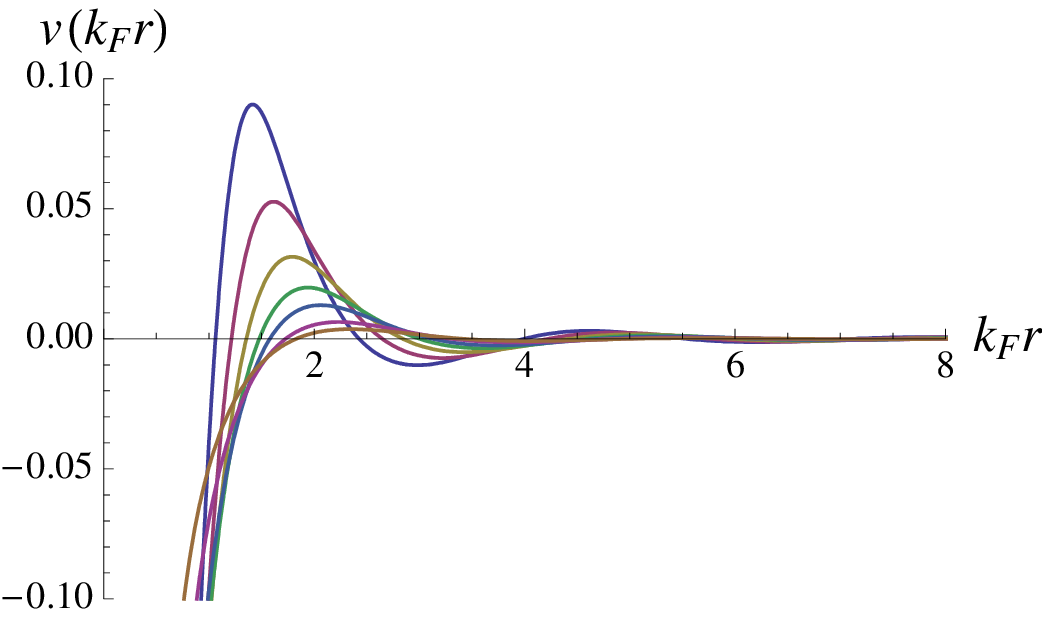}\hfill\hfill
 \includegraphics[width=0.46\textwidth,clip]{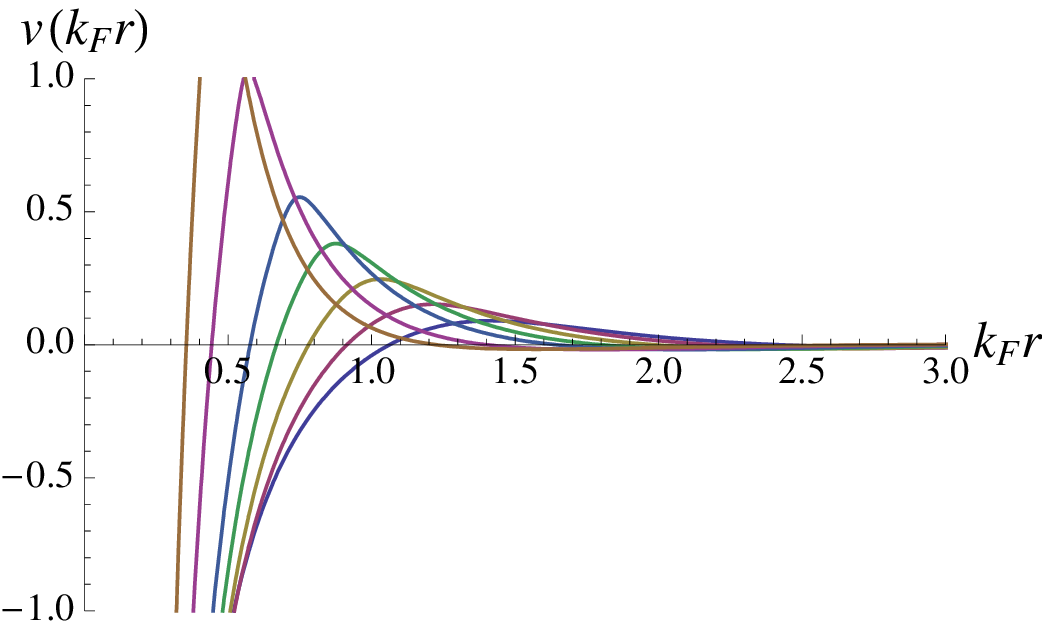}\hfill\hfill
 \caption{(Color online) Evolution of the interaction energy $v(\kF r)$
 as a function of the inverse $s$-wave scattering length $(a\kF)^{-1}$.
 Left panel: $(a\kF)^{-1}$ is varied from $-2$ (curve with the smallest
 amplitude) to $0$ (largest amplitude).  The tiny oscillatory behavior
 existing in the BCS regime grows toward the unitarity limit and makes a
 small hump at $r\simeq\kF^{-1}$.  Right panel: $(a\kF)^{-1}$ is varied
 from $0$ (curve with the smallest height) to $2$ (largest height).  The
 hump further grows in the BEC regime and develops a repulsive core at
 $r\simeq a$.  \label{fig:evolution}}
\end{figure*}

Now we fill the above-obtained energy levels with the light fermions.
The total energy of such a system is given by
\begin{equation}\label{eq:energy}
  E = -\frac{\kappa_+^{\,2}+\kappa_-^{\,2}}{2m}
 + \sum_{n=1}^{N}\frac{{k_n^+}^2+{k_n^-}^2}{2m},
\end{equation}
where $2N$ is the number of interacting light fermions in the continuum
states.  We compare the energy of the interacting system with that in
the noninteracting limit,
\begin{equation}
 E_\mathrm{free} = 
  \left.\sum_{n=1}^{N}\frac{{k_n^+}^2+{k_n^-}^2}{2m}\right|_{\delta_\pm=0},
\end{equation}
and define the energy reduction as 
$\Delta E(|\r|)\equiv E-E_\mathrm{free}$.  Then we take the
thermodynamic limit $N,R\to\infty$ with the Fermi momentum of the light
fermions $\kF\equiv N\pi/R$ fixed.  If we notice that the summation over
$n$ is dominated by $n\sim N\gg1$ and neglect small corrections of
$O(1/N)$, we obtain the following simple expression for the energy
reduction:
\begin{equation}\label{eq:reduction}
 \Delta E(|\r|) = -\frac{\kappa_+^{\,2}+\kappa_-^{\,2}}{2m}
  - \int_0^{\kF}\!dk\,k\,\frac{\delta_+(k)+\delta_-(k)}{m\,\pi}.
\end{equation}
We note that the phase shifts given in Eq.~(\ref{eq:delta}) are defined
to be in the range $0\leq\delta_\pm(k)\le\pi$.  The same result can be
obtained on a more general footing by using the functional integral
method (see Appendix~\ref{sec:functional}).

\subsection{Effective interaction between two heavy fermions}
When the two heavy fermions are infinitely separated, the energy
reduction approaches
\begin{equation}
 \Delta E(|\r|\to\infty) \to 2\mu_\mathrm{single},
\end{equation}
where $\mu_\mathrm{single}<0$ is the chemical potential of a single
heavy fermion immersed in the Fermi sea of light
fermions~\cite{Combescot:2007}:
\begin{equation}
 \mu_\mathrm{single} = -\frac{\kF^{\,2}}{2m}
  \frac{a\kF+\left[1+(a\kF)^2\right]
  \left[\frac\pi2+\arctan(a\kF)^{-1}\right]}{\pi(a\kF)^2}.
\end{equation}
The energy reduction with $2\mu_\mathrm{single}$ subtracted is regarded
as the interaction energy of the two heavy fermions:
\begin{equation}\label{eq:V(r)}
 \begin{split}
  V(|\r|) &\equiv E(|\r|) - E(|\r|\to\infty) \\
  &= \Delta E(|\r|)-2\mu_\mathrm{single}.
 \end{split}
\end{equation}
$V(|\r|)$ represents the effective interaction between two heavy
fermions induced by the interaction with the Fermi sea of light
fermions.  We note that the chemical potential of light fermions
$\mu_l\equiv\frac{\kF^{\,2}}{2m}$ is fixed here instead of their
particle number.

It is convenient to measure the interaction energy $V(r)$ in units of
the Fermi energy of light fermions and introduce a dimensionless
function $v(\kF r)$ as
\begin{equation}\label{eq:v(r)}
 V(r) \equiv \frac{\kF^{\,2}}{2m}v(\kF r).
\end{equation}
$v(\kF r)$ is a function of the separation between the heavy fermions
$\kF r$ and also the $s$-wave scattering length $a\kF$.  $v(\kF r)$ for
three typical values of $(a\kF)^{-1}=-5,\,0,\,5$ are plotted in
Fig.~\ref{fig:effective_potential}.  One can see the smooth evolution of
the effective interaction between the two heavy fermions as a function
of $(a\kF)^{-1}$ in Fig.~\ref{fig:evolution}.  In the BCS regime
$(a\kF)^{-1}\lesssim-1$, the effective interaction is attractive at
$r\lesssim|a|$ and has a tiny oscillatory behavior at $r\gtrsim|a|$.
This oscillatory behavior grows toward the unitarity limit
$(a\kF)^{-1}\approx0$ and makes a small hump at $r\simeq\kF^{-1}$ (see
the left panel of Fig.~\ref{fig:evolution}).  This hump further grows in
the BEC regime $(a\kF)^{-1}\gtrsim1$ and eventually develops a repulsive
core at $r\simeq a$ (see the right panel of Fig.~\ref{fig:evolution}).

It is worthwhile to compare our nonperturbative result (\ref{eq:V(r)})
with the perturbative calculation of the effective interaction in the
BCS limit $a\to-0$.  In the BCS limit $|a|\ll r,\kF^{-1}$, we have
$\kappa_\pm\to0$ and the phase shift
\begin{equation}
 \begin{split}
  \delta_\pm(k) &\to \frac{|a|}{r}\left[kr\pm\sin(kr)\right] \\
  & \pm \left(\frac{|a|}{r}\right)^2\cos(kr)
  \left[kr\pm\sin(kr)\right] + O(|a|^3).
 \end{split}
\end{equation}
Thus we can find the interaction energy to be
\begin{equation}
 v(\kF r) \to (|a|\kF)^2
  \frac{2\kF r\cos(2\kF r)-\sin(2\kF r)}{2\pi(\kF r)^4} + O(|a|^3),
\end{equation}
which has the same form as the Ruderman-Kittel-Kasuya-Yosida interaction
between magnetic impurities in a Fermi liquid~\cite{RKKY}.  Accordingly
the Fourier transform of the effective interaction $V(|\r|)$ in the BCS
limit becomes
\begin{align}\label{eq:V(p)}
 \tilde V(|\p|) &\equiv \int\!d\r\,e^{-i\p\cdot\r}V(|\r|) \\\notag
 &\to -\frac{|a|^2\kF}{2m}
 \left[2+\left(\frac{2\kF}{|\p|}-\frac{|\p|}{2\kF}\right)
 \ln\left|\frac{1+\frac{|\p|}{2\kF}}{1-\frac{|\p|}{2\kF}}\right|\right],
\end{align}
which correctly reproduces the effective interaction to the leading
order in $a$ obtained in Ref.~\cite{Bulgac:2006gh}~\footnote{For unequal
masses, the effective interaction in the BCS limit becomes
$\tilde V(|\p|)=-\left[\frac{2\pi(m+M)a}{mM}\right]^2\frac{m\kF}{2\pi^2}L\!\left(\frac{|\p|}{2\kF}\right)$,
where $L(z)=\frac12+\frac{1-z^2}{4z}\ln\!\left|\frac{1+z}{1-z}\right|$
is the static Lindhard function.  The limit of large mass ratio
$M/m\gg1$ coincides with Eq.~(\ref{eq:V(p)}).} as it should be.  Our
nonperturbative result (\ref{eq:V(r)}) can go beyond the perturbative
BCS regime to the strongly interacting unitary regime in a controlled
way by utilizing $M/m\gg1$.

Analytic expressions of $v(\kF r)$ in various limits are obtained in
Appendix~\ref{sec:limits}.

\section{$P$-wave projection of the effective interaction \label{sec:p-wave}}

\begin{figure*}[tp]\hfill
 \includegraphics[width=0.46\textwidth,clip]{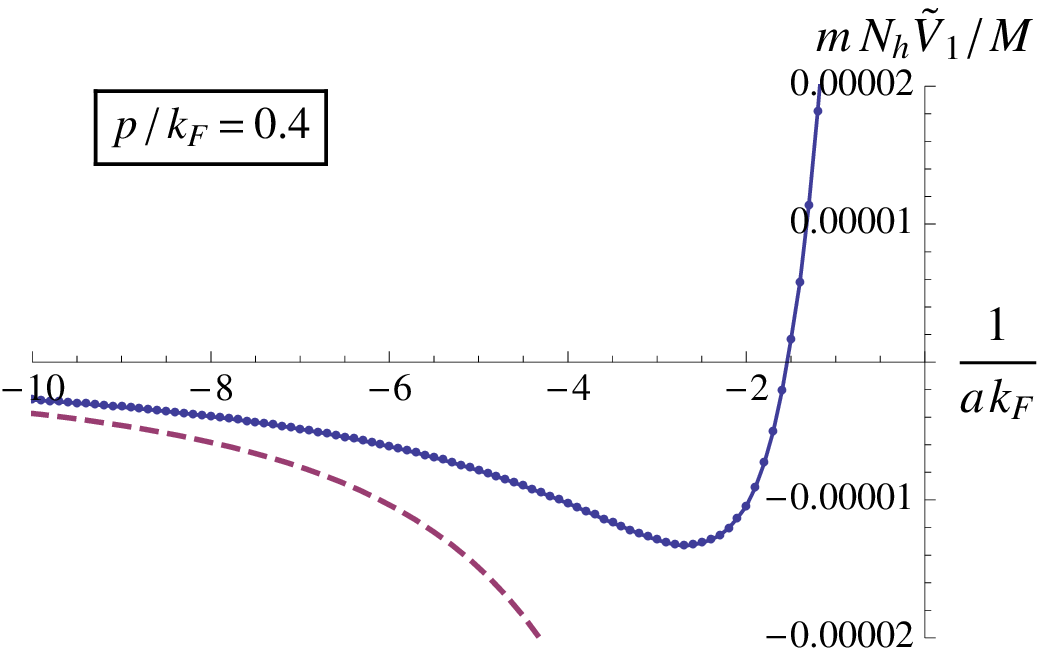}\hfill\hfill
 \includegraphics[width=0.46\textwidth,clip]{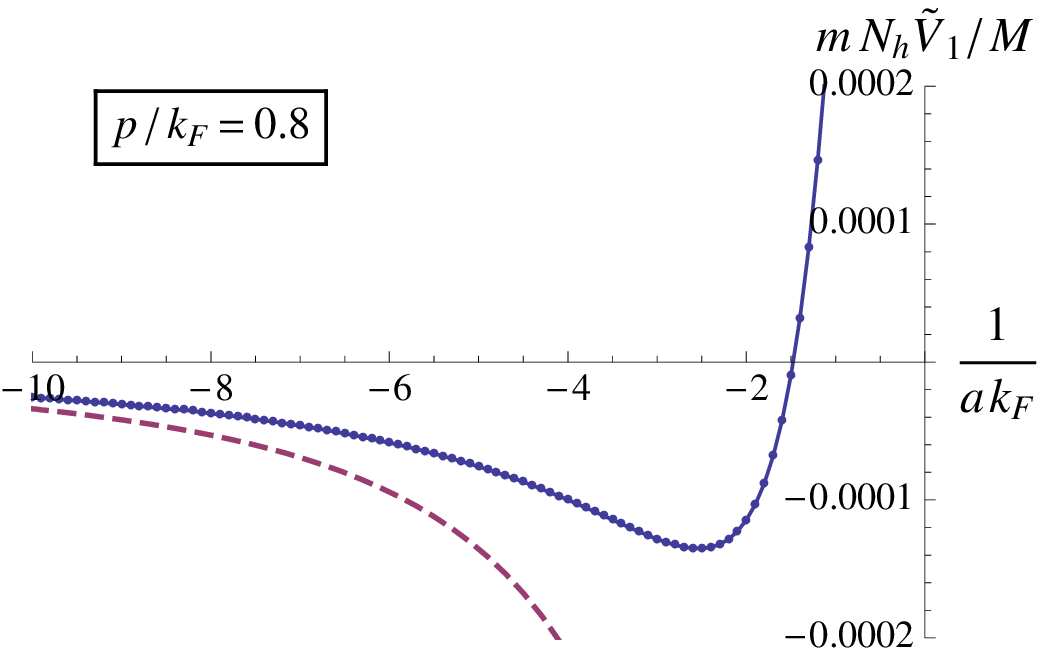}\hfill\hfill

 \bigskip\hfill
 \includegraphics[width=0.46\textwidth,clip]{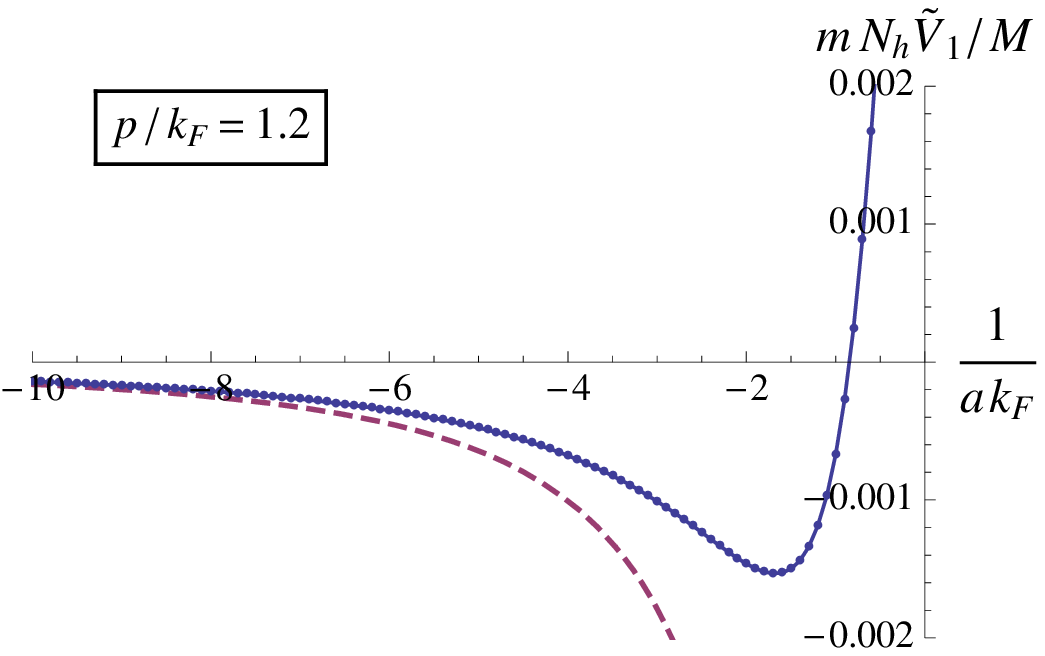}\hfill\hfill
 \includegraphics[width=0.46\textwidth,clip]{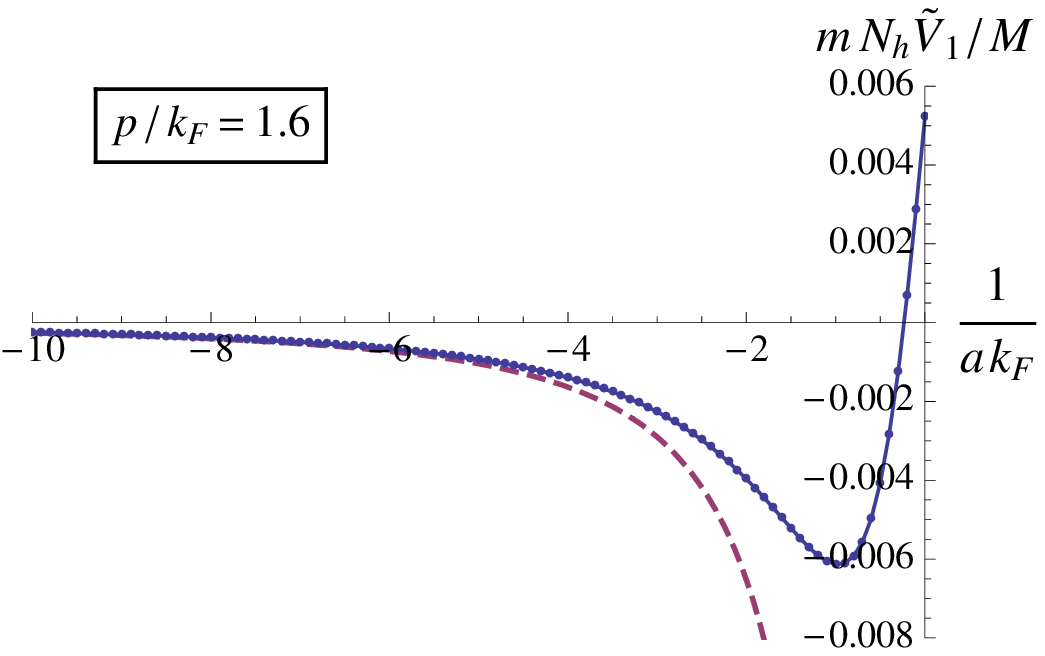}\hfill\hfill
 \caption{(Color online) $P$-wave effective interaction between two
 heavy fermions in three dimensions as a function of $(a\kF)^{-1}$.
 $\frac{m}{M}N_h\tilde V_{\ell=1}(p)$ in $d=3$ is plotted for
 $p/\kF=0.4$, $0.8$, $1.2$, and $1.6$.  The dashed curves are
 perturbative results at leading order valid in the BCS limit
 $(a\kF)^{-1}\to-\infty$.  \label{fig:3D_interaction}}
\end{figure*}

\subsection{Lessons from BCS theory}
The physics of heavy fermions immersed in the Fermi sea of light
fermions is described by the Hamiltonian
\begin{align}\label{eq:hamiltonian}
 H &= \int\!d\x\,\psi_h^\+(\x)
 \left(-\frac{\grad^2}{2M}-\mu_h\right)\psi_h(\x) \\\notag
 &\quad + \frac12\int\!d\x d\y\,\psi_h^\+(\x)\psi_h^\+(\y)
 V(|\x-\y|)\psi_h(\y)\psi_h(\x),
\end{align}
where $\mu_h$ is the chemical potential of heavy fermions measured from
$\mu_\mathrm{single}$.  Strictly speaking, the pairwise interaction
$V(|\r|)$ obtained in the previous section [Eq.~(\ref{eq:V(r)})] is
valid only if there are two heavy fermions in the system.  This is
because the Casimir interaction energy is not pairwise additive and also
because if heavy fermions have a finite density, it will perturb the
Fermi sea of light fermions and thus affect the effective interaction
$V(|\r|)$.  However, in the dilute limit $\sqrt{2M\mu_h}\ll\kF$, we
expect that the above Hamiltonian is a good approximation to the system
of heavy fermions because the probability to find a third heavy fermion
near the two heavy fermions becomes small in the dilute
system~\cite{Petrov:2007}.  Indeed we will confirm in
Appendix~\ref{sec:functional} that the Casimir interaction among more
than two heavy fermions at large separations can be evaluated quite
accurately as a sum of pairwise interactions between each of the two
heavy fermions.  This result also supports the use of our Hamiltonian
(\ref{eq:hamiltonian}) to describe the physics of dilute heavy fermions
immersed in the Fermi sea of light fermions.

Another issue to be addressed is the instability of the system.  When
the mass ratio $M/m$ exceeds the critical value 13.6 in pure
3D~\cite{Efimov:1972} or 6.35 in the 2D-3D
mixture~\cite{Nishida:2008kr}, the system will not be stable due to the
Efimov effect.  Therefore we need to assume the mass ratio to be large
but smaller than the above critical value.  For such a mass ratio, the
effective interaction $V(|\r|)$ obtained in the Born-Oppenheimer
approximation is no longer exact but can be considered as a good
approximation to the exact one.

One of predictions we can derive from the Hamiltonian
(\ref{eq:hamiltonian}) is a pairing between the heavy fermions.  The
pairing gap of heavy fermions $\Delta_\p$ is defined to be
\begin{align}
 & (2\pi)^d\delta(\k)\Delta_\p \\\notag
 &= \int\!\frac{d\q}{(2\pi)^d}\tilde V(|\p-\q|)
 \left\<\tilde\psi_h\!\left(\frac{\k}2-\q\right)
 \tilde\psi_h\!\left(\frac{\k}2+\q\right)\right\>,
\end{align}
where $\tilde\psi_h(\p)$ is the Fourier transform of $\psi_h(\x)$ and
$\tilde V(|\p-\q|)$ is the Fourier transform of $V(|\r|)$ [see
Eq.~(\ref{eq:V(p)})].  Because of the Fermi statistics of heavy
fermions, the pairing gap has to have an odd parity;
$\Delta_{-\p}=-\Delta_{\p}$.  The standard mean-field calculation leads
to the self-consistent gap equation
\begin{equation}\label{eq:gap_eq}
 \Delta_\p = -\int\!\frac{d\q}{(2\pi)^d}V(|\p-\q|)
  \frac{\Delta_\q}{2E_\q}\left[1-2\nF(E_\q)\right].
\end{equation}
Here $E_\p=\sqrt{\left(\frac{\p^2}{2M}-\mu_h\right)^2+|\Delta_\p|^2}$
is the quasiparticle energy and $\nF(E_\p)=1/\left(e^{E_\p/T}+1\right)$
is the Fermi-Dirac distribution function at temperature $T$.

When the coupling between different partial waves can be neglected, we
can solve the gap equation (\ref{eq:gap_eq}) easily by using the
weak-coupling approximation~\footnote{The weak-coupling approximation
even at unitarity is justified in the dilute limit;
$\sqrt{2M\mu_h}\ll\kF$.  From Eqs.~(\ref{eq:long-distance}) and
(\ref{eq:unitarity}), the interaction energy at the mean interparticle
distance of heavy fermions $r=(2M\mu_h)^{-1}\gg\kF^{-1}$ becomes
$V(r)\sim\frac{\kF^{\,2}}{2m}\frac1{(\kF r)^3}$ and is parametrically
small compared to the kinetic energy $K(r)\sim\frac{1}{2M r^2}$.}.
For a given orbital angular momentum $\ell$ (odd integer), the pairing
gap and the critical temperature are given by~\cite{Anderson:1961}
\begin{equation}\label{eq:Tc}
 |\Delta_\p| \sim \Tc \sim \mu_h
  \exp\!\left(\frac1{N_h\tilde V_\ell}\right).
\end{equation}
Here $N_h$ is the density of states of heavy fermions at the Fermi
surface and $\tilde V_\ell$ is the partial-wave projection of the
effective interaction $\tilde V(|\p-\q|)$ with
$|\p|=|\q|=\sqrt{2M\mu_h}$ fixed on the Fermi surface of heavy fermions
and assumed to be attractive $\tilde V_\ell<0$.  Because the lowest
partial wave for identical fermions is $\ell=1$, the dominant pairing is
expected to occur in the $p$-wave channel.  With having in mind the
application to the pairing of heavy fermions, we compute 
$N_h\tilde V_\ell$ as a function of the $s$-wave scattering length
$a\kF$ from Eq.~(\ref{eq:V(r)}).  Below we consider both cases where the
heavy fermions are in three dimensions ($d=3$) and where they are
confined in two dimensions ($d=2$).

\begin{figure*}[tp]\hfill
 \includegraphics[width=0.46\textwidth,clip]{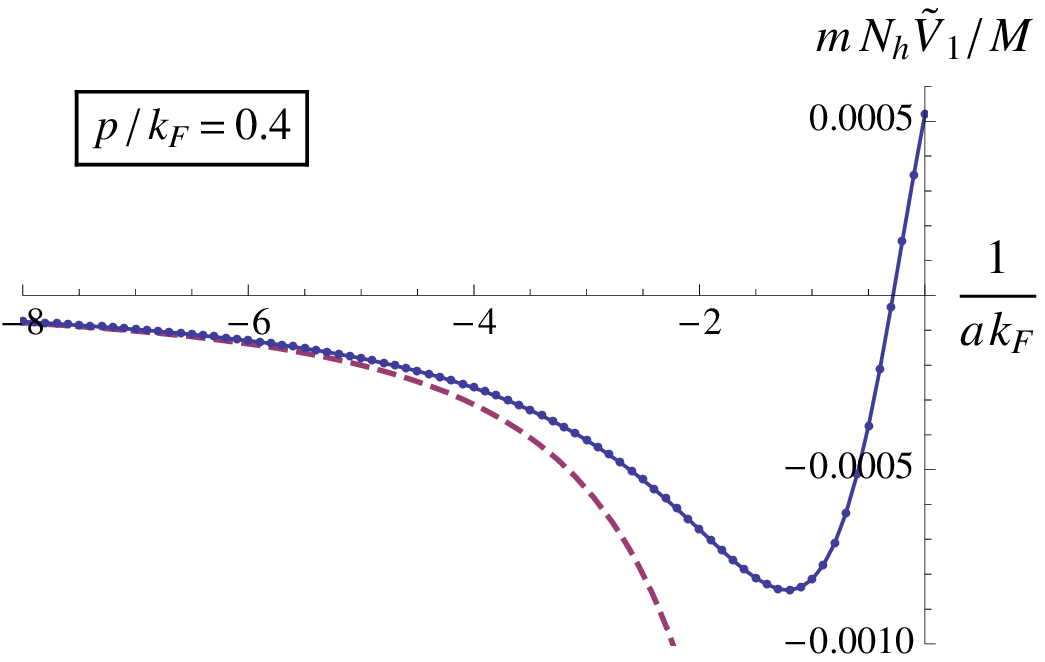}\hfill\hfill
 \includegraphics[width=0.46\textwidth,clip]{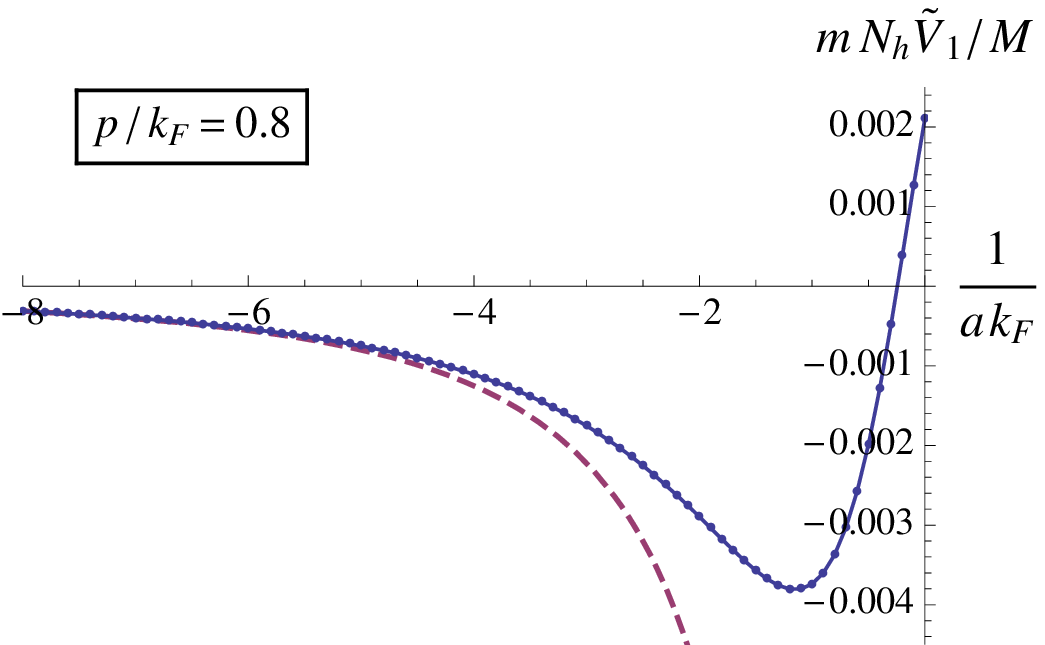}\hfill\hfill

 \bigskip\hfill
 \includegraphics[width=0.46\textwidth,clip]{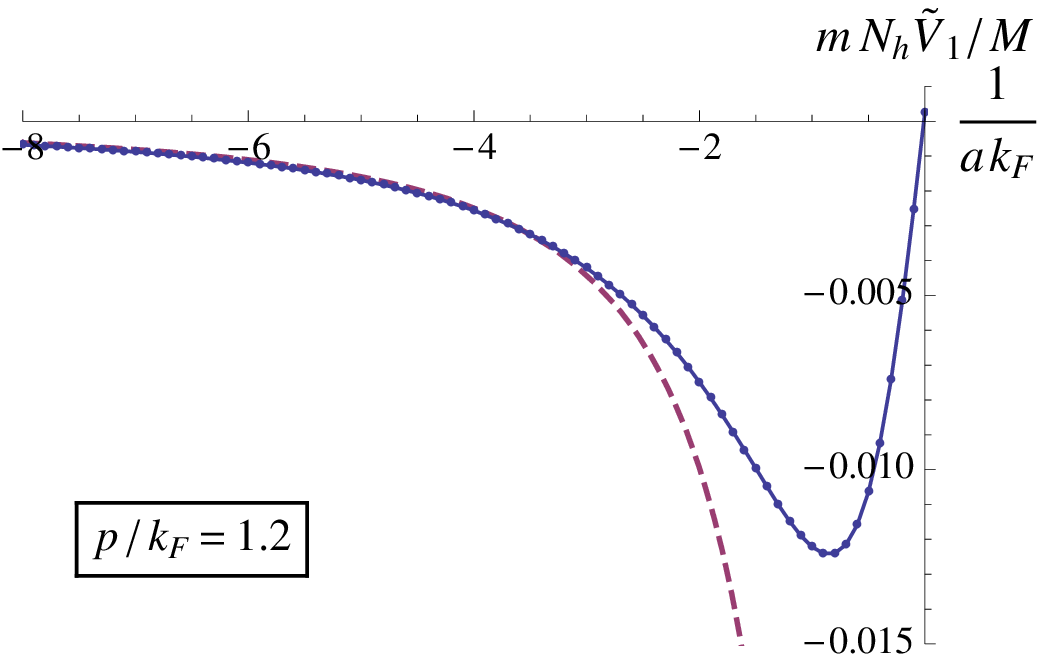}\hfill\hfill
 \includegraphics[width=0.46\textwidth,clip]{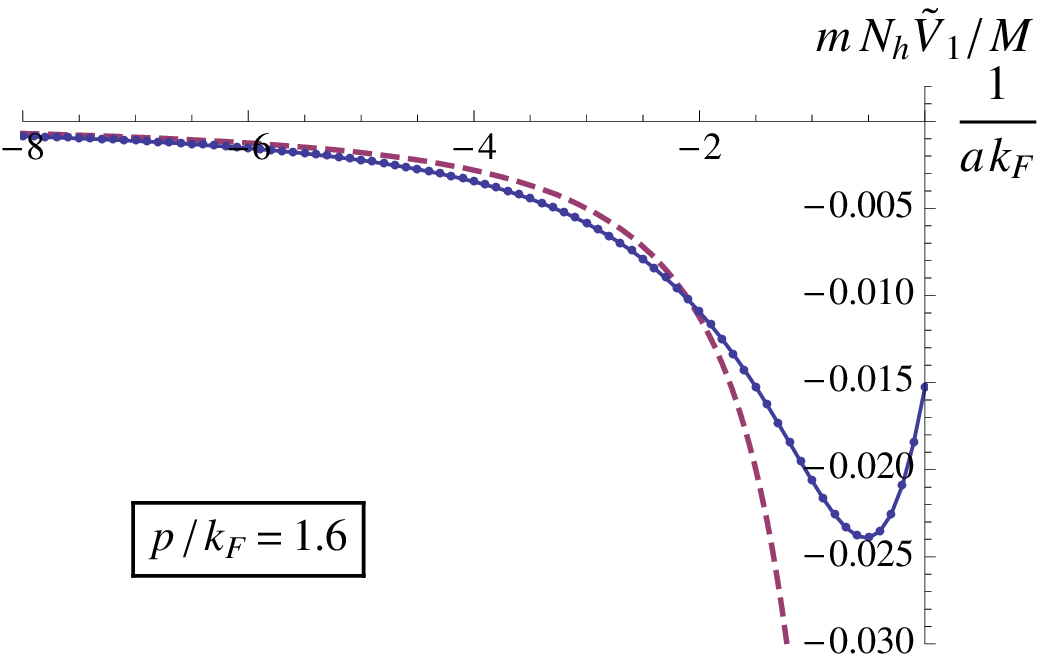}\hfill\hfill
 \caption{(Color online) $P$-wave effective interaction between two
 heavy fermions confined in two dimensions as a function of
 $(a\kF)^{-1}$.  $\frac{m}{M}N_h\tilde V_{\ell=1}(p)$ in $d=2$ is
 plotted for $p/\kF=0.4$, $0.8$, $1.2$, and $1.6$.  The dashed curves
 are perturbative results at leading order valid in the BCS limit
 $(a\kF)^{-1}\to-\infty$.  \label{fig:2D_interaction}}
\end{figure*}

\subsection{Heavy fermions in $d=3$}
When the heavy fermions are in three dimensions $d=3$, the partial-wave
projection of the effective interaction $\tilde V(|\p-\q|)$ with
$|\p|=|\q|\equiv p$ fixed is given by
\begin{equation}
 \tilde V_\ell(p) = \frac12\int_{-1}^1d\cos\theta
  \,P_\ell(\cos\theta)\,\tilde V(|\p-\q|),
\end{equation}
where $\cos\theta=\hat\p\cdot\hat\q$.  Multiplying it by the density of
states of heavy fermions $N_h(p)=\frac{Mp}{2\pi^2}$, we obtain the
dimensionless function representing the effective interaction between
two heavy fermions for the given partial wave $\ell$:
\begin{equation}
 \begin{split}
  N_h\tilde V_\ell(p)
  &= \frac{M}{2\pi m}\left(\frac{\kF}{p}\right)^2
  \int_{-1}^{1}d\cos\theta\,P_\ell(\cos\theta) \\
  &\quad \times 
  \int_0^\infty\!dz\,\frac{z\sin\!\left(z\sqrt{2-2\cos\theta}\right)}
  {\sqrt{2-2\cos\theta}}\,v\!\left(\frac{\kF z}{p}\right).
 \end{split}
\end{equation}
When the projected effective interaction with $p=\sqrt{2M\mu_h}$ being
on the Fermi surface of heavy fermions is attractive
$N_h\tilde V_\ell<0$, the pairing of the heavy fermions is expected to
occur with the pairing gap and the critical temperature given in
Eq.~(\ref{eq:Tc}).

Figure~\ref{fig:3D_interaction} shows $\frac{m}{M}N_h\tilde V_\ell(p)$ in
the $p$-wave channel $\ell=1$ as a function of $(a\kF)^{-1}$ for four
values of $p/\kF=0.4,\,0.8,\,1.2,\,1.6$.  We note that what is plotted
is not the $p$-wave effective interaction
$N_h\tilde V_\ell(p)\propto\frac{M}{m}$ itself but
$\frac{m}{M}N_h\tilde V_\ell(p)$ that is independent of the mass ratio
$M/m\gg1$.  We can see that the $p$-wave effective interaction is
attractive in the BCS limit $(a\kF)^{-1}\ll-1$ being consistent with the
perturbative prediction~\cite{Bulgac:2006gh}.  However, it turns out
that the perturbative result has only a small range of validity.  Our
result shows that the $p$-wave attraction is generally weaker than the
extrapolation of the perturbative calculation and, in particular, the
$p$-wave effective interaction becomes a strong repulsion near the
unitarity limit.  This may be understood because of the repulsive hump
of $v(\kF r)$ developed near the unitarity limit (see
Figs.~\ref{fig:effective_potential} and \ref{fig:evolution}).  We also
find that the $p$-wave attraction is stronger for the larger value of
$p/\kF$ because the contribution from the attractive part of
$v(\kF r)\to-c^2/(\kF r)^2$ at short distance $r\to0$ [see
Eq.~(\ref{eq:short-distance})] becomes more significant for larger
$p/\kF$.  The maximum attraction for each value of
$p/\kF=0.4,\,0.8,\,1.2,\,1.6$ is achieved around
$(a\kF)^{-1}\approx-2.7,\,-2.5,\,-1.7,-1.0$, respectively.  At such a
value of the $s$-wave scattering length, the maximal $p$-wave pairing of
heavy minority fermions immersed in the Fermi sea of light fermions is
possible while the pairing gap and the critical temperature would be
very small because of the numerically weak attraction
$N_h\tilde V_{\ell=1}(p)\ll1$.  One can perform the same analysis for
the $f$-wave channel ($\ell=3$) but will find even weaker attraction.

\subsection{Heavy fermions in $d=2$}
When the heavy fermions are confined in two dimensions $d=2$, the
partial-wave projection of the effective interaction $\tilde V(|\p-\q|)$
with $|\p|=|\q|\equiv p$ fixed is given by
\begin{equation}
 \tilde V_\ell(p) = \frac1\pi\int_0^\pi\!d\theta
  \,\cos(\ell\theta)\,\tilde V(|\p-\q|),
\end{equation}
where $\cos\theta=\hat\p\cdot\hat\q$.  Multiplying it by the density of
states of heavy fermions $N_h(p)=\frac{M}{2\pi}$, we obtain the
dimensionless function representing the effective interaction between
two heavy fermions for the given partial wave $\ell$:
\begin{equation}
 \begin{split}
  N_h\tilde V_\ell(p)
  &= \frac{M}{2\pi m}\left(\frac{\kF}{p}\right)^2
  \int_0^\pi\!d\theta\,\cos(\ell\theta) \\
  &\quad \times 
  \int_0^\infty\!dz\,z\,J_0\!\left(z\sqrt{2-2\cos\theta}\right)
  v\!\left(\frac{\kF z}{p}\right).
 \end{split}
\end{equation}
In the case of $d=2$, the scattering length $a$ should be regarded as
the effective scattering length $a_\mathrm{eff}$ introduced in
Ref.~\cite{Nishida:2008kr}.

Figure~\ref{fig:2D_interaction} shows $\frac{m}{M}N_h\tilde V_\ell(p)$
in the $p$-wave channel $\ell=1$ as a function of $(a\kF)^{-1}$ for four
values of $p/\kF=0.4,\,0.8,\,1.2,\,1.6$.  One can see the similar
behavior to the case of $d=3$.  Again the $p$-wave effective interaction
is attractive in the BCS limit $(a\kF)^{-1}\ll-1$ being consistent with
the perturbative prediction~\cite{Nishida:2008gk}.  Compared to the case
of $d=3$, we find that the perturbative result has a wider range of
validity.  In the case of $d=2$, the $p$-wave attraction is found to be
stronger than that in $d=3$ for the same value of $p/\kF$ and the
$p$-wave effective interaction becomes only a weak repulsion near the
unitarity limit.  This will be understandable because the contribution
from the attractive part of $v(\kF r)\to-c^2/(\kF r)^2$ at short
distance $r\to0$ [see Eq.~(\ref{eq:short-distance})] is less suppressed
by the phase space factor and thus more significant in $d=2$.  Our
result also shows that the $p$-wave attraction is stronger for the
larger value of $p/\kF$ and the maximum attraction for each value of
$p/\kF=0.4,\,0.8,\,1.2,\,1.6$ is achieved around
$(a\kF)^{-1}\approx-1.2,\,-1.2,\,-0.85,\,-0.52$, respectively.  At such
a value of the $s$-wave scattering length, the maximal $p$-wave pairing
of heavy minority fermions in two dimensions immersed in the
three-dimensional Fermi sea of light fermions is possible and the
pairing gap and the critical temperature will be larger than those in
the case of $d=3$.

\section{Summary and concluding remarks \label{sec:summary}}
We have investigated a two-species Fermi gas with a large mass ratio
interacting by an interspecies short-range interaction.  Using the
Born-Oppenheimer approximation, we determined the interaction energy
of two heavy fermions immersed in the Fermi sea of light fermions, which
is an analog of the Casimir force, as a function of the $s$-wave
scattering length.  We showed that the $p$-wave projection of the
effective interaction is attractive in the BCS limit being consistent
with the perturbative prediction, while it turns out to be repulsive
near the unitarity limit.  We found that the $p$-wave attraction reaches
its maximum between the BCS and unitarity limits, where the maximal but
still weak $p$-wave pairing of heavy minority fermions is possible.

We also investigated the case where the heavy fermions are confined in
two dimensions corresponding to the two-species Fermi gas in the 2D-3D
mixed dimensions~\cite{Nishida:2008kr,Nishida:2008gk}.  Because the
$p$-wave attraction between the heavy fermions in two dimensions was
found to be stronger than that in three dimensions, we expect the larger
$p$-wave pairing gap of heavy minority fermions.  Such a $p$-wave
pairing in two dimensions is especially interesting because the
resulting system has a potential application to topological quantum
computation using vortices with non-Abelian
statistics~\cite{Read:2000,Tewari:2007}.

Although our controlled analysis is restricted to the dilute heavy
fermions in the limit of large mass ratio, our results indeed shed light
on the phase diagram of asymmetric Fermi gases with unequal densities
and masses (and spatial dimensions), in particular, in the
strongly interacting unitary regime.  Our results also have a direct
relevance to the recently realized Fermi-Fermi mixture of
${}^{40}\mathrm{K}$ and ${}^6\mathrm{Li}$ because of their large mass
ratio~\cite{Taglieber:2008,Wille:2008,Voigt:2008}.  It will be an
important future problem to study corrections to our results when the
restrictions of the large mass ratio and the dilute limit of heavy
fermions are relaxed.

\acknowledgments
The author thanks M.~M.~Forbes, R.~Jaffe, S.~Tan, and M.~Zwierlein for
discussions.  This work was supported by MIT Pappalardo Fellowships in
Physics.

\appendix
\section{Casimir interaction among heavy fermions from
functional integral method \label{sec:functional}}

\subsection{Derivation of Casimir interaction}
Here we derive the Casimir interaction among heavy fermions immersed in
the Fermi sea of light fermions using the functional integral
method~\cite{Recati:2005}.  The action that describes the light fermions
with the chemical potential $\mu_l$ interacting with the heavy fermions
fixed at positions $\x_i$ is
\begin{equation}
 \begin{split}
  S &= \int\!d\tau d\x\,
  \psi_l^\+(\tau,\x)\left(\d_\tau-\frac{\grad^2}{2m}-\mu_l\right)\psi_l(\tau,\x) \\
  &\quad - g_0\sum_i\int\!d\tau\,\psi_l^\+(\tau,\x_i)\psi_l(\tau,\x_i),
 \end{split}
\end{equation}
where $\tau$ is an imaginary time.  The partition function
$Z=e^{-\beta\Omega}$ is given by
\begin{equation}
 Z = \int\mathcal{D}\psi_l\mathcal{D}\psi_l^\+e^{-S}.
\end{equation}
We first insert the following identity into the integrand:
\begin{equation}
 1 = \prod_i\int\mathcal{D}\eta_i\mathcal{D}\bar\eta_i
  \,\delta[\psi_l(\tau,\x_i)-\eta_i(\tau)]
  \delta[\psi_l^\+(\tau,\x_i)-\bar\eta_i(\tau)]
\end{equation}
and then exponentiate the delta functions by introducing auxiliary
fields,
\begin{align}
  & 1 = \prod_i\int\mathcal{D}\eta_i\mathcal{D}\bar\eta_i
  \mathcal{D}\alpha_i\mathcal{D}\bar\alpha_i \\\notag
  & \times e^{i\int\!d\tau\bar\alpha_i(\tau)[\psi_l(\tau,\x_i)-\eta_i(\tau)]
  + i\int\!d\tau[\psi_l^\+(\tau,\x_i)-\bar\eta_i(\tau)]\alpha_i(\tau)}.
\end{align}
Now the partition function can be written as
\begin{equation}
 Z = \int\mathcal{D}\psi_l\mathcal{D}\psi_l^\+
  \prod_i\mathcal{D}\eta_i\mathcal{D}\bar\eta_i
  \mathcal{D}\alpha_i\mathcal{D}\bar\alpha_i\,e^{-S'},
\end{equation}
where the action becomes
\begin{equation}
 \begin{split}
  S' &= \int\!d\tau d\x\,\psi_l^\+(\tau,\x)
  \left(\d_\tau-\frac{\grad^2}{2m}-\mu_l\right)\psi_l(\tau,\x) \\
  &\quad - g_0\sum_i\int\!d\tau\,\bar\eta_i(\tau)\eta_i(\tau) \\
  &\quad - i\sum_i\int\!d\tau\,
  \bar\alpha_i(\tau)[\psi_l(\tau,\x_i)-\eta_i(\tau)] \\
  &\quad - i\sum_i\int\!d\tau\,
  [\psi_l^\+(\tau,\x_i)-\bar\eta_i(\tau)]\alpha_i(\tau).
 \end{split}
\end{equation}

We can easily integrate out $\psi_l$ and $\psi_l^\+$ fields and $\eta$
and $\bar\eta$ fields to lead to the partition function
\begin{equation}
 Z = Z_0\int\prod_i\mathcal{D}\alpha_i\mathcal{D}\bar\alpha_i\,e^{-S''},
\end{equation}
where $Z_0=e^{-\beta\Omega_\mathrm{free}}$ is a partition function of
noninteracting light fermions and the action $S''$ in the momentum space
becomes
\begin{equation}
 \begin{split}
  S'' &= \sum_{i,j}\int\frac{d\omega d\p}{(2\pi)^4}\,
  \bar\alpha_i(\omega)\frac{e^{i\p\cdot(\x_i-\x_j)}}
  {-i\omega+\frac{\p^2}{2m}-\mu_l}\alpha_j(\omega) \\
  &\quad - \frac1{g_0}\sum_i\int\frac{d\omega}{2\pi}\,
  \bar\alpha_i(\omega)\alpha_i(\omega).
 \end{split}
\end{equation}
Finally, by integrating out $\alpha$ and $\bar\alpha$ fields, we obtain
the following expression for the partition function:
\begin{equation}
 Z = Z_0\exp\!
  \left[\beta\int\frac{d\omega}{2\pi}\ln\det M_{ij}(i\omega)\right],
\end{equation}
where $M_{ij}$ is the scattering matrix given by
\begin{align}
 M_{ij} &= -\frac{\delta_{ij}}{g_0}
 + \int\!\frac{d\p}{(2\pi)^3}\frac{e^{i\p\cdot(\x_i-\x_j)}}
 {-i\omega+\frac{\p^2}{2m}-\mu_l} \\\notag
 &= 
 \begin{cases}
  \displaystyle
  \,\frac{m}{2\pi}\left(\frac1a-\sqrt{-2mi\omega-2m\mu_l}\right)
  & \text{for}\quad i=j \medskip\\
  \displaystyle
  \,\frac{m}{2\pi|\x_i-\x_j|}e^{-\sqrt{-2mi\omega-2m\mu_l}|\x_i-\x_j|}
  & \text{for}\quad i\neq j.
 \end{cases}
\end{align}
Here we introduced the $s$-wave scattering length $a$ through
\begin{equation}
 -\frac1{g_0} + \int\!\frac{d\p}{(2\pi)^3}\frac{2m}{\p^2}
  = \frac{m}{2\pi a}.
\end{equation}
Therefore the reduction in the grand potential $\Omega$ compared to that
in the noninteracting limit $\Omega_\mathrm{free}$ is
\begin{equation}
 \Delta\Omega(\{\x_i\}) = -\int\frac{d\omega}{2\pi}\ln\det M_{ij}(i\omega).
\end{equation}
By subtracting $\Delta\Omega$ in which all heavy fermions are infinitely
separated, the interaction energy of the heavy fermions becomes
\begin{equation}\label{eq:casimir}
 V(\{\x_i\}) = -\int\frac{d\omega}{2\pi}
  \ln\frac{\det M_{ij}(i\omega)}{\det M_{ij}^\infty(i\omega)},
\end{equation}
where $M_{ij}^\infty$ is a diagonal scattering matrix composed of
$M_{ii}$.  This is the generalization of $V(|\r|)$ in
Eq.~(\ref{eq:V(r)}) to a general number of heavy fermions fixed at
positions $\x_i$.

\begin{figure*}[tp]
 \includegraphics[width=0.33333\textwidth,clip]{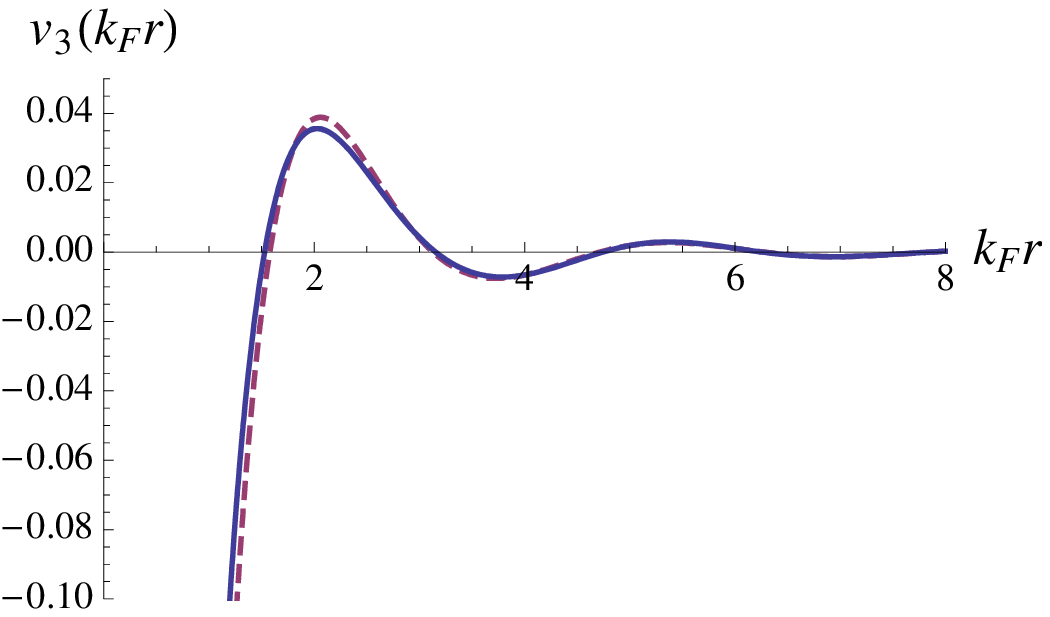}\hfill
 \includegraphics[width=0.33333\textwidth,clip]{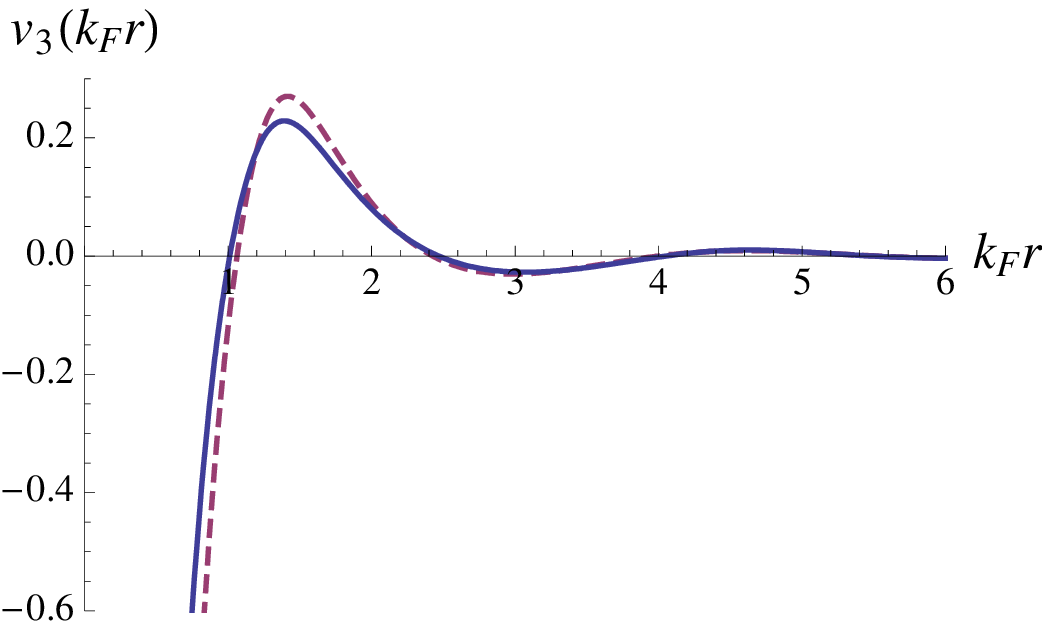}\hfill
 \includegraphics[width=0.33333\textwidth,clip]{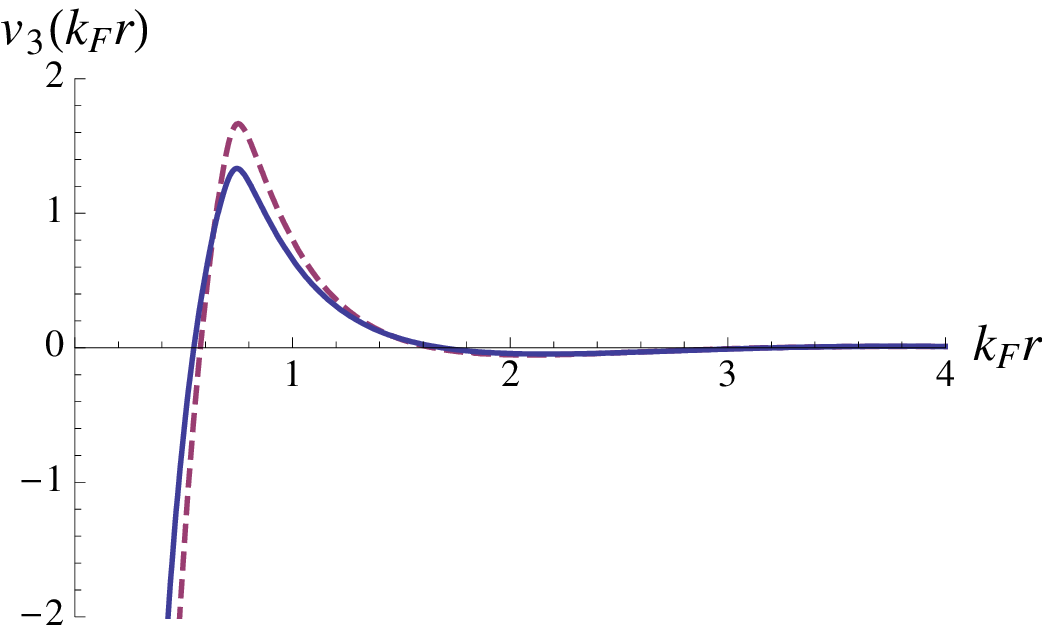}
 \caption{(Color online) Interaction energy of three heavy fermions
 fixed with the same separations $r$.  $v_3(\kF r)$ as a function of
 $\kF r$ is plotted for $(a\kF)^{-1}=-1$ (left), $0$ (middle), and $1$
 (right).  The dashed curves represent the sum of pairwise interaction
 energies $3v_2(\kF r)$.  \label{fig:three}}
\end{figure*}

\begin{figure*}[tp]
 \includegraphics[width=0.33333\textwidth,clip]{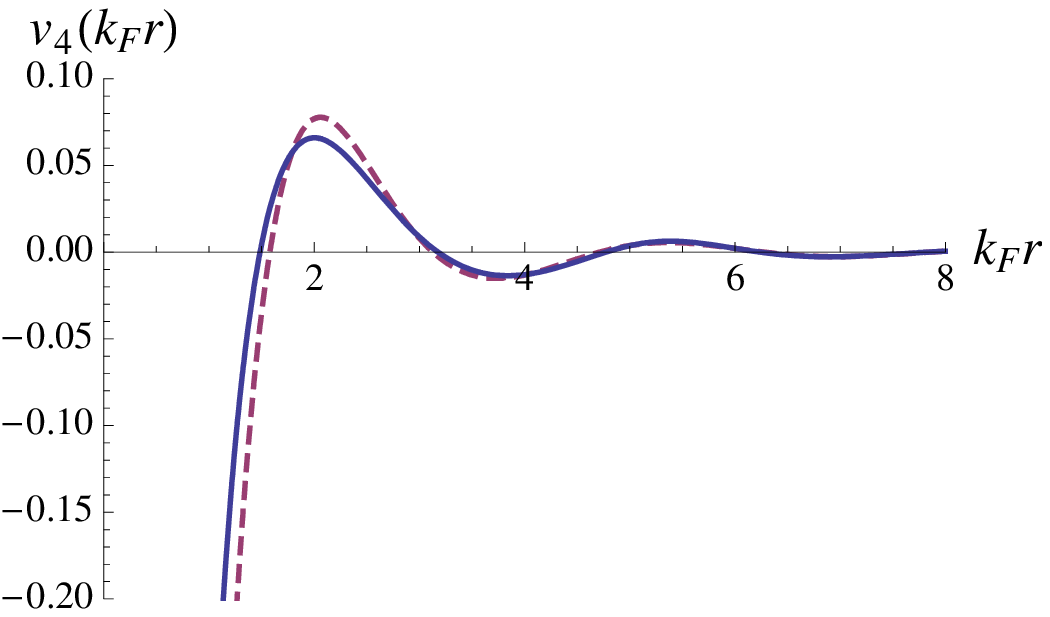}\hfill
 \includegraphics[width=0.33333\textwidth,clip]{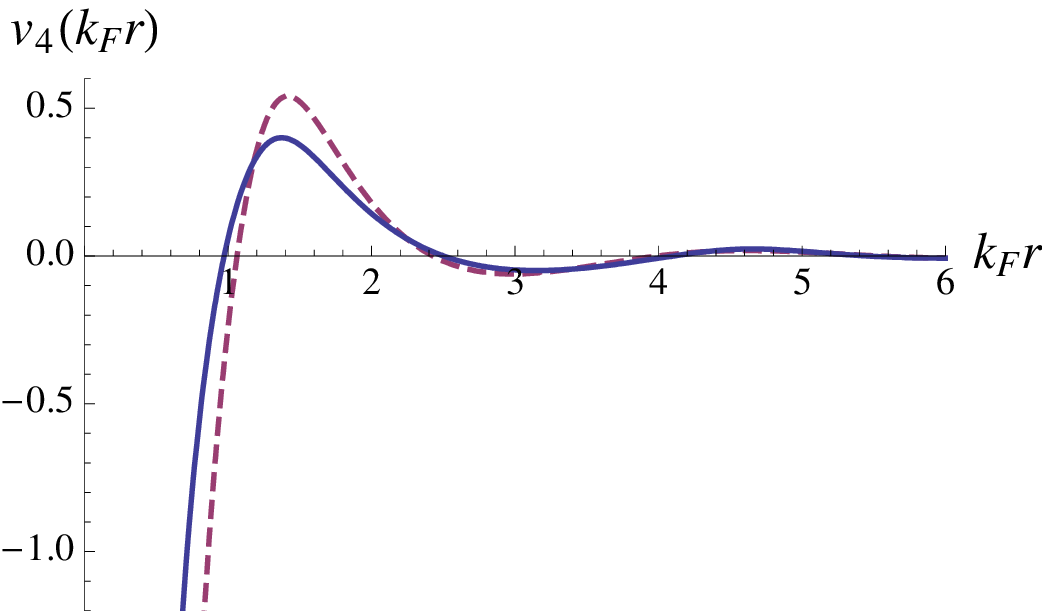}\hfill
 \includegraphics[width=0.33333\textwidth,clip]{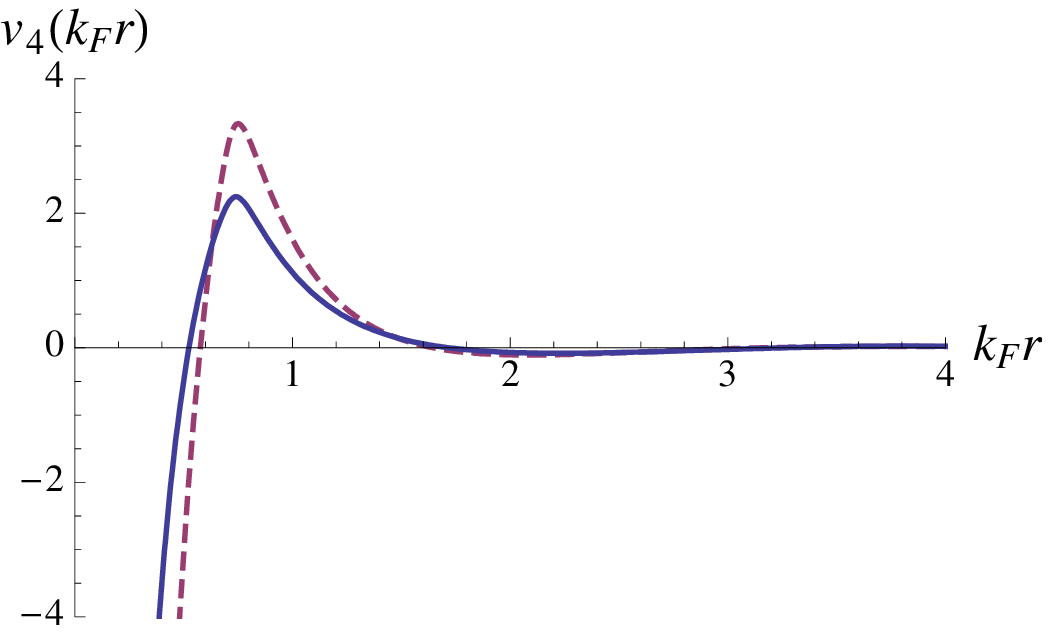}
 \caption{(Color online) Interaction energy of four heavy fermions fixed
 with the same separations $r$.  $v_4(\kF r)$ as a function of $\kF r$
 is plotted for $(a\kF)^{-1}=-1$ (left), $0$ (middle), and $1$ (right).
 The dashed curves represent the sum of pairwise interaction energies
 $6v_2(\kF r)$.  \label{fig:four}}
\end{figure*}

\subsection{Casimir interaction between two heavy fermions}
It is worthwhile to reproduce the result obtained in
Sec.~\ref{sec:casimir} in the case of two heavy fermions.  By deforming
the path of the integration over $i\nu\equiv i\omega+\mu_l$ into
\begin{equation}
 \int_{-i\infty+\mu_l}^{i\infty+\mu_l}
  \to \int_{-\infty-i0^+}^{\mu_l} + \int_{\mu_l}^{-\infty+i0^+},
\end{equation}
we obtain the following expression for the grand potential reduction:
\begin{equation}\label{eq:grand_potential}
 \begin{split}
  \Delta\Omega(|\r|)
  &= i\int\frac{d\nu}{2\pi}\ln\!\left(\frac1a-\sqrt{-2m\nu}
  + \frac{e^{-\sqrt{-2m\nu}|\r|}}{|\r|}\right) \\
  &\ + i\int\frac{d\nu}{2\pi}\ln\!\left(\frac1a-\sqrt{-2m\nu}
  - \frac{e^{-\sqrt{-2m\nu}|\r|}}{|\r|}\right)
 \end{split}
\end{equation}
with $\r=\x_1-\x_2$.  We separate the integral into the contributions
from bound states $\mathrm{Re}[\nu]<0$ and continuum states
$\mathrm{Re}[\nu]>0$.  For the bound state contribution, we pick up the
singularity in the integrand (\ref{eq:grand_potential}) as
\begin{align}
 & i\int_{\mathrm{Re}[\nu]<0}
 \frac{d\nu}{2\pi}\ln\!\left(\frac1a-\sqrt{-2m\nu}
 \pm \frac{e^{-\sqrt{-2m\nu}|\r|}}{|\r|}\right) \notag\\
 &= \nu_{\pm} + \mathrm{const},
\end{align}
where $\nu_\pm<0$ is the binding energy satisfying
\begin{equation}
 \frac1a-\sqrt{-2m\nu_{\pm}}
  \pm \frac{e^{-\sqrt{-2m\nu_{\pm}}|\r|}}{|\r|} = 0.
\end{equation}
For the continuum state contribution, defining the phase shift by
\begin{align}
 & \pi - \delta_\pm(\nu) \\\notag
 &= \arg\!\left(\frac1a-\sqrt{-2m\nu-i0^+}
 \pm \frac{e^{-\sqrt{-2m\nu-i0^+}|\r|}}{|\r|}\right),
\end{align}
the integral in Eq.~(\ref{eq:grand_potential}) can be written as
\begin{align}
 & i\int_{\mathrm{Re}[\nu]>0}
 \frac{d\nu}{2\pi}\ln\!\left(\frac1a-\sqrt{-2m\nu}
 \pm \frac{e^{-\sqrt{-2m\nu}|\r|}}{|\r|}\right) \notag\\
 &= \mu_l - \int_0^{\mu_l}\!d\nu\,\frac{\delta_\pm(\nu)}\pi.
\end{align}
Therefore, after dropping the unimportant constants, we find that the
grand potential reduction in the case of two heavy fermions is given by
\begin{equation}
 \Delta\Omega(|\r|) = \nu_+ + \nu_-
  - \int_0^{\mu_l}\!d\nu\,
  \frac{\delta_+(\nu)+\delta_-(\nu)}\pi.
\end{equation}
This result is equivalent to $\Delta E(|\r|)$ in
Eq.~(\ref{eq:reduction}) with $\nu_\pm=-\frac{\kappa_\pm^{\,2}}{2m}$,
$\nu=\frac{k^2}{2m}$, and $\mu_l=\frac{\kF^{\,2}}{2m}$ and thus provides
the same interaction energy $V(|\r|)$ as Eq.~(\ref{eq:V(r)}).

\subsection{Casimir interaction among three and four heavy fermions}
We now evaluate the Casimir interaction energy (\ref{eq:casimir}) among
three and four heavy fermions fixed with the same separations
$|\x_i-\x_j|\equiv r$.  We measure the interaction energy in units of
the Fermi energy of light fermions
$V_N(r)\equiv\frac{\kF^{\,2}}{2m}v_N(\kF r)$ with the subscript $N$
indicating the number of heavy fermions.  The dimensionless functions
$v_N(\kF r)$ with three typical values of $(a\kF)^{-1}=-1,\,0,\,1$ are
plotted as functions of $\kF r$ in Fig.~\ref{fig:three} for $N=3$ and in
Fig.~\ref{fig:four} for $N=4$, together with the sum of pairwise
interaction energies $\frac{N(N-1)}{2}v_2(\kF r)$ for comparison.

We can see that the Casimir interaction among more than two heavy
fermions can be reproduced quite accurately by the sum of pairwise
interactions between each of the two heavy fermions, in particular, at
long distances, although the behaviors at short distances are slightly
overestimated.  A similar observation has been made in
Ref.~\cite{Bulgac:2001np} in the system of hardcore spheres immersed in
a background Fermi sea.  These results support the use of the
Hamiltonian (\ref{eq:hamiltonian}) only with the pairwise interaction to
describe the physics of dilute heavy fermions immersed in the Fermi sea
of light fermions.

\section{Interaction energy in various limits \label{sec:limits}}
Here we evaluate the effective interaction between two heavy fermions
$v(\kF r)$ in Eq.~(\ref{eq:v(r)}) in various limits where analytic
expressions are available.

At short distance $r\ll a,\kF^{-1}$, we obtain
\begin{align}\label{eq:short-distance}
 v(\kF r) &\to -\frac{c^2}{(\kF r)^2}
 - \frac{2c}{1+c}\frac1{a\kF^{\,2}r}
 - \frac{1+c+c^2}{(1+c)^3}\frac1{a^2\kF^{\,2}} \notag\\
 &\quad + \frac{2a\kF+\left[1+(a\kF)^2\right]
 \left[\pi+2\arctan(a\kF)^{-1}\right]}{\pi(a\kF)^2} \notag\\
 &\quad - 1 + O(r),
\end{align}
where $c=0.567143$ is a solution to $c=e^{-c}$.

On the other hand, at long distance $r\gg a,\kF^{-1}$, we obtain
\begin{align}\label{eq:long-distance}
  &v(\kF r) \\ &\to \frac{2(a\kF)^3\sin(2\kF r)
  -(a\kF)^2\left[(a\kF)^2-1\right]\cos(2\kF r)}
  {\pi\left[(a\kF)^2+1\right]^2(\kF r)^3} \notag\\
  &\quad - \frac{4(a\kF)^3\left[(a\kF)^2-1\right]\cos(2\kF r)}
  {2\pi\left[(a\kF)^2+1\right]^3(\kF r)^4} \notag\\
  &\quad - \frac{(a\kF)^2\left[(a\kF)^4-6(a\kF)^2+1\right]\sin(2\kF r)}
  {2\pi\left[(a\kF)^2+1\right]^3(\kF r)^4}
 + O(r^{-5}). \notag
\end{align}
In particular, in the unitarity limit $a\kF\to\infty$, we find
\begin{align}\label{eq:unitarity}
 v(\kF r) &\to - \frac{\cos(2\kF r)}{\pi(\kF r)^3}
 - \frac{\sin(2\kF r)}{2\pi(\kF r)^4} \\\notag
 &\quad + \frac{2\cos(2\kF r)+\cos(4\kF r)}{4\pi(\kF r)^5}
 + O\!\left[(\kF r)^{-6}\right].
\end{align}

In the BCS or BEC limit $|a|\ll r,\kF^{-1}$, we obtain
\begin{align}
 & v(\kF r) \to (a\kF)^2
 \frac{2\kF r\cos(2\kF r)-\sin(2\kF r)}{2\pi(\kF r)^4} \notag\\
 &\quad + (a\kF)^3 
 \frac{2\kF r\cos(2\kF r)+\left[2(\kF r)^2-1\right]\sin(2\kF r)}
 {\pi(\kF r)^5} \notag\\
 &\quad + O(a^4).
\end{align}


\begin{thebibliography}{99}

\bibitem{Ketterle:2008}
  W.~Ketterle and M.~W.~Zwierlein,
  {\em Proceedings of the International School of Physics ``Enrico Fermi,''\/}
  Varenna (IOS Press, Amsterdam, 2008),
  arXiv:0801.2500, and references therein.

\bibitem{review_theory}
  For recent theoretical reviews, see
%
  I.~Bloch, J.~Dalibard, and W.~Zwerger,
  Rev.\ Mod.\ Phys.\ {\bf 80}, 885 (2008);
%
  S.~Giorgini, L.~P.~Pitaevskii, and S.~Stringari,
  Rev.\ Mod.\ Phys.\ {\bf 80}, 1215 (2008).

\bibitem{Zwierlein:2005}
  M.~W.~Zwierlein, A.~Schirotzek, C.~H.~Schunck, and W.~Ketterle,
  Science {\bf 311}, 492 (2006).

\bibitem{Partridge:2005}
  G.~B.~Partridge, W.~Li, R.~I.~Kamar, Y.~Liao, and R.~G.~Hulet,
  Science {\bf 311}, 503 (2006).

\bibitem{Zwierlein:2006}
  M.~W.~Zwierlein, C.~H.~Schunck, A.~Schirotzek, and W.~Ketterle,
  Nature (London) {\bf 442}, 54 (2006).

\bibitem{Shin:2006}
  Y.~Shin, M.~W.~Zwierlein, C.~H.~Schunck, A.~Schirotzek, and W.~Ketterle,
  Phys.\ Rev.\ Lett.\ {\bf 97}, 030401 (2006).

\bibitem{Partridge:2006}
  G.~B.~Partridge, W.~Li, Y.~A.~Liao, R.~G.~Hulet, M.~Haque, and H.~T.~C.~Stoof,
  Phys.\ Rev.\ Lett.\ {\bf 97}, 190407 (2006).

\bibitem{Taglieber:2008}
  M.~Taglieber, A.-C.~Voigt, T.~Aoki, T.~W.~H\"ansch, and K.~Dieckmann,
  Phys.\ Rev.\ Lett.\ {\bf 100}, 010401 (2008).

\bibitem{Wille:2008}
  E.~Wille, F.~M.~Spiegelhalder, G.~Kerner, D.~Naik, A.~Trenkwalder,
  G.~Hendl, F.~Schreck, R.~Grimm, T.~G.~Tiecke, J.~T.~M.~Walraven,
  S.~J.~J.~M.~F.~Kokkelmans, E.~Tiesinga, and P.~S.~Julienne, 
  Phys.\ Rev.\ Lett.\ {\bf 100}, 053201 (2008).

\bibitem{Voigt:2008}
  A.-C.~Voigt, M.~Taglieber, L.~Costa, T.~Aoki, W.~Wieser, T.~W.~H\"ansch,
  and K.~Dieckmann,
  arXiv:0810.1306.

\bibitem{Alford:2007xm}
  For a recent review, see
  M.~G.~Alford, A.~Schmitt, K.~Rajagopal, and T.~Schafer,
  Rev.\ Mod.\ Phys.\ {\bf 80}, 1455 (2008).

\bibitem{Bulgac:2006gh}
  A.~Bulgac, M.~M.~Forbes, and A.~Schwenk,
  Phys.\ Rev.\ Lett.\ {\bf 97}, 020402 (2006).

\bibitem{Batrouni:2008}
  G.~G.~Batrouni, M.~H.~Huntley, V.~G.~Rousseau, and R.~T.~Scalettar,
  Phys.\ Rev.\ Lett.\ {\bf 100}, 116405 (2008).

\bibitem{Casula:2008}
  M.~Casula, D.~M.~Ceperley, and E.~J.~Mueller,
  Phys.\ Rev.\ A {\bf 78}, 033607 (2008).

\bibitem{Batrouni:2009}
  G.~G.~Batrouni, M.~J.~Wolak, F.~Hebert, and V.~G.~Rousseau,
  arXiv:0809.4549.

\bibitem{Casimir:1948dh}
  H.~B.~G.~Casimir,
  Proc.\ K.\ Ned.\ Akad.\ Wet.\ {\bf 51}, 793 (1948).

\bibitem{Bulgac:2001np}
  A.~Bulgac and A.~Wirzba,
  Phys.\ Rev.\ Lett.\ {\bf 87}, 120404 (2001).

\bibitem{Nishida:2008kr}
  Y.~Nishida and S.~Tan,
  Phys.\ Rev.\ Lett.\ {\bf 101}, 170401 (2008).

\bibitem{Nishida:2008gk}
  Y.~Nishida,
  arXiv:0810.1321, to be published in Ann.\ Phys.\ (N.Y.).

\bibitem{Combescot:2007}
  R.~Combescot, A.~Recati, C.~Lobo, and F.~Chevy,
  Phys.\ Rev.\ Lett.\ {\bf 98}, 180402 (2007).

\bibitem{RKKY}
  M.~A.~Ruderman and C.~Kittel,
  Phys.\ Rev.\ {\bf 96}, 99 (1954).

  T.~Kasuya,
  Prog.\ Theor.\ Phys.\ {\bf 16}, 45 (1956).

  K.~Yosida,
  Phys.\ Rev.\ {\bf 106}, 893 (1957).

\bibitem{Petrov:2007}
  D.~S.~Petrov, G.~E.~Astrakharchik, D.~J.~Papoular, C.~Salomon,
  and G.~V.~Shlyapnikov,
  Phys.\ Rev.\ Lett.\ {\bf 99}, 130407 (2007).

\bibitem{Efimov:1972}
  V.~Efimov, 
  Sov.\ Phys.\ JETP Lett.\ {\bf 16}, 34 (1972);
  Nucl.\ Phys.\ {\bf A210}, 157 (1973).

\bibitem{Anderson:1961}
  P.~W.~Anderson and P.~Morel,
  Phys.\ Rev.\ {\bf 123}, 1911 (1961).

\bibitem{Read:2000}
  N.~Read and D.~Green,
  Phys.\ Rev.\ B {\bf 61}, 10267 (2000).

\bibitem{Tewari:2007}
  S.~Tewari, S.~Das~Sarma, C.~Nayak, C.~Zhang, and P.~Zoller,
  Phys.\ Rev.\ Lett.\ {\bf 98}, 010506 (2007).

\bibitem{Recati:2005}
  A.~Recati, J.~N.~Fuchs, C.~S.~Pe\c{c}a, and W.~Zwerger,
  Phys.\ Rev.\ A {\bf 72}, 023616 (2005).

\end{thebibliography}
\end{document}